\documentclass[aps,twocolumn,superscriptaddress,showpacs,showkeys]{revtex4}
\usepackage{graphics,graphicx,dcolumn,bm,fleqn,epic,eepic,float}
\usepackage{amssymb,amsmath,multirow,rotate,color,float}
\usepackage{times}
\usepackage{latexsym}
\usepackage{wasysym}

\begin{document}

\title{Vegetation against dune mobility}

\author{Orencio Dur\'an}
\affiliation{Institute for Computational Physics, 
             Universit\"at Stuttgart, Pfaffenwaldring 27, 
             D-70569 Stuttgart, Germany}
\author{Hans J.~Herrmann}
\affiliation{Institute for Computational Physics, 
             Universit\"at Stuttgart, Pfaffenwaldring 27, 
             D-70569 Stuttgart, Germany}
\affiliation{Departamento de F\'{\i}sica, Universidade Federal do
             Cear\'a, 60451-970 Fortaleza, Brazil}

\begin{abstract}
Vegetation is the most common and most reliable stabilizer of loose soil or sand. This ancient technique is for the first time cast into a set of equations of motion describing the competition between aeolian sand transport and vegetation growth. Our set of equations is then applied to study quantitatively the transition between barchans and parabolic dunes driven by the dimensionless fixation index $\theta$ which is the ratio between dune characteristic erosion rate and vegetation growth velocity. We find a fixation index $\theta_c$ below which the dunes are stabilized characterized by scaling laws.
\end{abstract}


\maketitle

Since pharaonic times mobile sand has been stabilized through plantations while conversely fields and woodland have been devastated by wind-driven erosion and coverage of sand. This ancient fight between vegetation growth and aeolian surface mobility has evidently enormous impact on the economy of semi-arid regions, on coastal management and on global ecosystems. While empirical techniques have been systematically improved since dune fields in Aquitaine were fixed during the reign of Louis XIV and an entire specialty has developed in agronomy\cite{gilad,hardenberg,tsoar90} most approaches are just based on trial and error and there is still an astonishing lack of mathematical description and of quantitative predictability. It is the aim of this Letter to propose for the first time a set of differential equations of motion describing the aeolian transport on vegetated granular surfaces including the growth and destruction of the plants. Vegetation hinders sand mobility but also becomes its victim through erosion of the roots and coverage by sand. We will also show that these new equations can be used to calculate the morphological transition between barchans and parabolic dunes.

Migrating crescent dunes occupy about half of the total desert and coastal area under uni-modal wind condition \cite{wasson}, and they exist either as isolated barchan dunes in places with sparse sand or forming barchanoid ridges where sand is abundant (Fig.~1, top left). These migrating dunes can be deactivated by the invasion of vegetation, as seen in Fig.~1 (top), provided that there is a certain amount of rainfall and weak human interference \cite{tsoar90,tsoar05}. When the vegetation cover grows, barchans apparentely undergo a transformation into parabolic dunes with arms pointing upwind and partly colonized by plants (Fig.~1, top center-right), a metamorphosis that has been seen as the first step of dune inactivation \cite{hack,anthonsen,muckersie,tsoar02}. 

\begin{figure*}
\includegraphics[width=0.85\textwidth]{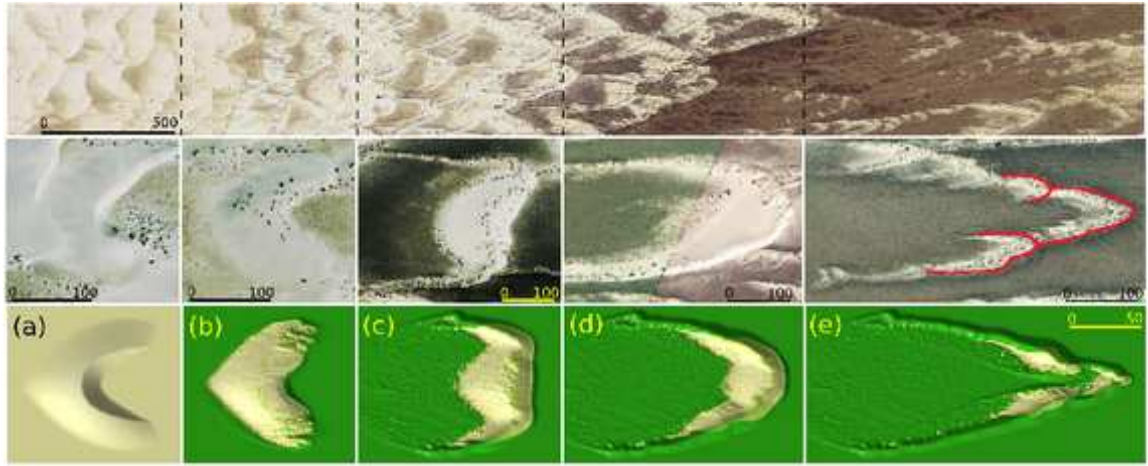}
\caption{Deactivation of migrating dunes under the influence of vegetation. On top, a dune field in White Sand, New Mexico, that shows barchanoid ridges on the left, where vegetation is absent, developing towards a mixture of active and inactive parabolic dunes on the right (wind blows from left to right). Dark green regions indicate abundance of vegetation. This suggests a transition between both types of dunes when the vegetation cover increases. This transition is illustrated with various dune types found in the White Sand dune field (pictures in the middle), reinforcing the idea of their common evolution from a crescent dune. Satellite images taken from GoogleEarth. Below, same transition obtained by the numerical solution of a model that accounts for the coupling between sand transport and vegetation, with fixation index $\theta=0.22$. The vegetation cover is represented in green (grey).}
\end{figure*}

The complex coupling between vegetation and aeolian sand transport involves two different time scales related, on one hand, to vegetation growth and erosion and deposition processes that change the surface, and, on the other hand, to sand transport and wind flow. A significant change in vegetation or in sand surface can happen within some hours or even days. In contrast, the time scale of wind flow changes and saltation process is of the order of seconds and therefore orders of magnitude faster. This separation of time scales leads to an enormous simplification because it decouples the different processes. Hence, we can use stationary solutions for the wind surface shear stress $\vec \tau$ and for the resulting sand flux $\vec q$. 

The evolution of the sand surface and the vegetation that grows over it remains however coupled. Since plants locally slow down the wind, they can inhibit sand erosion as well as enhance sand accretion. This local slowdown of the wind shear stress represents the main dynamical effect of the vegetation over the wind field and hence on the sand flux. The vegetation acts as a roughness element that absorbs part of the momentum transferred to the soil by the wind. As a result, the total surface shear stress $\tau\equiv |\vec \tau|$ can be divided into two components, $\tau_v$ acting on the vegetation and $\tau_s$ on the sand grains. When plants are randomly distributed and the effective shelter area for one plant is assumed to be proportional to its frontal area, the absorbed shear stress $\tau_v$ is proportional to the vegetation frontal area density $\lambda$ times the undisturbed shear $\tau_s$ \cite{raupach}. Therefore, the fraction $\tau_s$ of total stress acting on sand grains is reduced to
\begin{equation}
\label{taus}
\tau_s = \tau / (1+m \beta \lambda) \, ,
\end{equation}
where $\beta$ is the ratio of plant to surface drag coefficients and $\lambda = \rho_v/\sigma$, where $\rho_v$ is the vegetation cover density and $\sigma$ the ratio of plant basal to frontal area. $m$ is a model parameter \cite{raupach,wyatt}. If the plant frontal area density $\lambda$ is zero there is no shear stress reduction. Otherwise, the reduction depends on the frontal area and on the plant drag coefficient encoded in $\beta$.

The time evolution of the sand surface height $h$ is calculated using the conservation of mass,
\begin{equation}
\label{h}
\partial h/\partial t = -\nabla \cdot \vec q(\tau_s)\, ,
\end{equation}
defining the erosion rate $\partial h/\partial t$ in terms of the sand flux $\vec q$ resulting from a wind surface shear stress $\tau_s$ that includes the vegetation feedback.

Additionally, the sand dynamics affects the growth of plants, since non-cohesive sand is severely eroded by strong winds denuding the roots and increasing the evaporation from deep layers. Although desert plants can resist quite severe conditions, sand erosion often kills them \cite{danin,hesp,bowers}. We assume therefore that plants under sand erosion or deposition need extra time to adapt to the surface change and the vegetation growth rate is delayed by the net erosion rate. Otherwise, they can grow up to a maximum height $H_v$ during a characteristic growth time $t_v$ \cite{richards}
\begin{equation}
\label{hv}
dh_v/dt = V_v\,(1 - h_v/H_v) - \left|\partial h/\partial t\right|
\end{equation}
where $h_v$ is the actual plant height, related with the vegetation cover density introduced in Eq.~\ref{taus} by $\rho_v=(h_v/H_v)^2$. Besides, the vegetation growth velocity $V_v\equiv H_v/t_v$ contains the information of climatic and local conditions that enhance or inhibit the growth process \cite{danin,hesp,bowers}.

This model for the interaction between plants and moving sand is closed by using an explicit formulation for the sand transport over a complex terrain \cite{kroy,sauermann,veit}. Analytical calculations of the flow over a gentle hill yield an analytical expression for the topographically induced perturbation $\delta \vec \tau$ of the surface shear stress $\vec \tau_0$ on a flat bed \cite{weng}, therefore 
\begin{equation}
\label{tau}
\vec \tau(h) = \vec \tau_0 + \tau_0 \delta \vec \tau (h),
\end{equation}
where $\tau_0 \equiv |\vec \tau_0|$. Whereas the stationary sand flux $\vec q$ is given by the logistic equation 
\begin{equation}
\label{q}
\nabla \cdot \vec q = q\,(1-q/q_s)/l_s \, ,
\end{equation}
where $q\equiv |\vec q|$, $q_s\equiv |\vec q_s|$, $\vec q_s(\tau)$ is the saturated sand flux, i.e. the maximum sand flux carried by the wind with a surface shear $\tau$, and $l_s(\tau)$ is the characteristic length of saturation transients, called saturation length.

Given an initial sand surface, a surface shear stress $\tau_0$ on a flat bed and a sand influx, the time evolution of the complete system is determined by the consecutive integration of the equations for $\tau(h)$ (Eq.~\ref{tau}), $\tau_s$ (Eq.~\ref{taus}), $\vec q$ (Eq.~\ref{q}), $h$ (Eq.~\ref{h}) and $h_v$ (Eq.~\ref{hv}). 

Figures 1a-1e show the effect of growing vegetation on the mobility of a barchan that evolves under a constant uni-directional wind (Fig.~1a). From Eq.~\ref{hv} plants will grow wherever $V_v > |\partial h/\partial t|$ even if under erosion conditions ($\partial h/\partial t < 0$) they can not germinate due to the mobile sand surface. Therefore, we introduce an heterogeneous seeding by allowing plants to germinate only where no erosion occurs, establishing a competition for their survival with the mobile sand. Such a scenario could for instance be the consequence of the cessation of human activity \cite{tsoar02}, an increase of the annual precipitation or a reduction of the wind strength in a dune field \cite{anthonsen,gaylord}, all of them stimulating conditions for vegetation growth. 

Plants first invade locations with small enough erosion or deposition, like the horns, the crest and the surroundings of the dune. This vegetation reduces the strength of the wind and traps the sand which accumulates mainly where sand flux is highest, i.e. at the crest. Consequently, the sand cannot reach the lee side anymore and deposits on the crest (Fig.~1b). This deposition kills the vegetation again. Thus the erosion/deposition rate at the crest is again lowered. This provides a cyclic mechanism for the vegetation growth on the crest.

As a result, the central part of the dune moves forward and two marginal ridges are left behind at the horns (Fig.~1c) in a process that leads to the stretching of the windward side and the formation of a parabolic dune (Fig.~1d). Finally, vegetation overcomes sand erosion even in the central part and the migration velocity of the parabolic dune dramatically drops about 80\%  indicating its inactivation (Fig.~1e). This process agrees well with a recent conceptual model based on field observations \cite{tsoar02}.

The transformation of barchans into parabolic dunes is determined by the initial barchan volume $V$, the undisturbed saturated sand flux $Q= |\vec q_s(\tau_0)|$, which encodes the wind strength in the flat bed, and the vegetation growth rate $V_v$. In fact, the stabilization process depends on the competition between sand transport and vegetation growth expressed by Eq.(\ref{hv}). This competition can be quantified by a dimensionless control parameter $\theta$, which we call fixation index, defined as the ratio between the characteristic erosion rate on the initial barchan dune, which scales as $Q$ to the volume defined characteristic length $V^{1/3}$ ratio (Eq.~\ref{h}), and the vegetation growth rate,
\begin{equation}
\label{theta}
\theta \equiv  Q/(V^{1/3}V_v) \, .
\end{equation}

After performing simulations for different $V$, $Q$ and $V_v$, we find a common fixation index $\theta_c \approx 0.5$ beyond which vegetation fails to complete the inversion of the barchan dune. As long as $\theta > \theta_c$ barchan dunes remain mobile, otherwise they are deactivated by plants.

\begin{figure}[htb]
\includegraphics[width=0.5\textwidth]{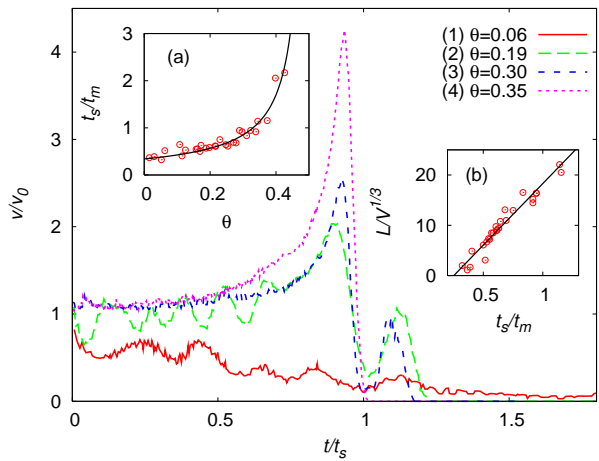}
\caption{Evolution of the normalized dune velocity ($v/v_0$) for simulations with initial condition given by the parameters $V$ (x$10^3$ m$^3$), $Q$ (m$^2$/yr) and $V_v$ (m/yr): (1) 200, 286 and 78, (2) 200, 286 and 26, (3) 44, 529 and 52, and (4) 200, 529 and 26. They represent fixation indexes $\theta$: 0.06 (1), 0.19 (2), 0.30 (3) and 0.35 (4), respectively. Inset (a): the normalized dune inactivation time $t_s/t_m$ diverges when $\theta \to \theta_c$ (solid line). Inset (b): the normalized final parabolic dune length $L/V^{1/3}$ is proportional to $t_s/t_m$. Dots indicate simulation results.}
\end{figure}

This morphologic transition at $\theta_c$ is typified by the evolution of the normalized velocity $v/v_0$ of vegetated dunes for different initial conditions (some of them are shown in Fig.~2), where $v_0$ is the initial barchan velocity. They reveal a general pattern characterized for medium and high fixation indexes by a speedup preceeding a sharp decrease of the dune velocity at $t=t_s$ when inactivation occurs (the explicit condition for inactivation reads $v/v_0 < 0.2$. At $t<t_s$ the barchan evolves toward an active parabolic dune (Fig.~1, (a)-(d)) that becomes inactive for larger times ($t>t_s$) (Fig.~1, (e)). However, for fixation index larger than $\theta_c$ barchans remain mobile and thus $t_s$ diverges. 

Since the dune velocity $v$ is proportional to the inverse of its height \cite{bagnold}, the sharp speedup in Fig.~2 is a consequence of the height decrease due to the redistribution of sand in the elongated dune. This redistribution process is enhanced at high $\theta$ when vegetation is progressively weaker against sand erosion. At small fixation indexes however, vegetation overcomes sand deposition and dune movement is determined by the vegetation dying rate on the crest which results in a slower dune motion. In this regime the dune velocity oscillates due to the cyclic mechanism that controls the vegetation growth on the crest. The same mechanism is responsable for the short reactivation of the inactive parabolic dune represented by the occasional second peak in the dune velocity for $t>t_s$ (Fig.~2).

From the analysis of the inactivation time $t_s$ as function of $V$, $Q$ and $V_v$, we find that $t_s$ scales with a characteristic dune migration time $t_m\equiv V^{2/3}/Q$ and obeys a power law in $\theta$ (Fig.~2, inset(a))
\begin{equation}
\label{ts}
t_s \approx 0.17\,t_m/(\theta_c-\theta) \,, 
\end{equation}
which discloses a transition between an inactive parabolic dune and a barchan dune remaining active. 

\begin{figure}[htb]
\includegraphics[width=0.4\textwidth]{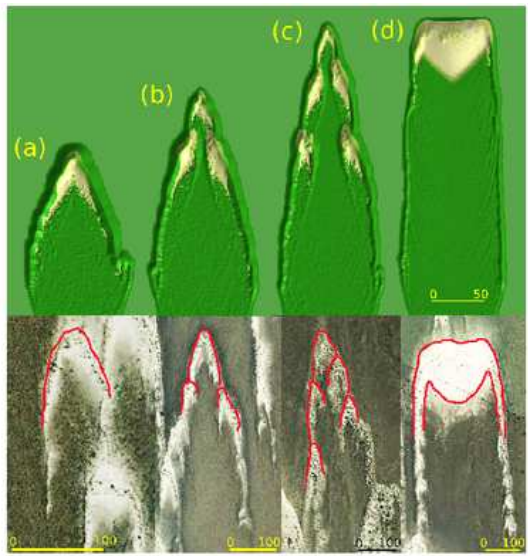}
\caption{Examples of parabolic dune shapes at fixation indexes $\theta$: (a) 0.16, (b) 0.22 and (c) 0.27. (d) is an example of an intermediate state at $\theta=0.38$ where the dune still is barchan-like, in sharp contrast to the rest. Each figure is compared with real examples from White Sand, New Mexico. Note the similarities in the contour lines market in red. Satellite images taken from GoogleEarth.}
\end{figure}

The normalized final length of the parabolic dune $L/V^{1/3}$ also diverges near the transition point. The parabolic length depends linearly on the normalized inactivation time, $L/V^{1/3} \approx a\,t_s/t_m - r$, (Fig.~2, inset (b)) where $a\approx 24.5$ and $r\approx 6.2$, and thus scales as $L/V^{1/3}\sim (\theta_c-\theta)^{-1}$. Note that in the limit $\theta=0$, the inactivation time is still proportional to $t_m$ ($t_s\approx 0.34\,t_m$) and the dune moves a distance $L\approx 2\,V^{1/3}$. This is consequence of the heterogeneous seeding assumed before. Since plants can not grow in the dune windward side where erosion occurs, all this sand is stabilized only after crossing the crest. Thus, the dune moves a distance equivalent to its windward side length $\sim 2\,V^{1/3}$.

Figure 3 shows some examples of the $L$ dependence on $\theta$. For small $\theta$, barchan dunes are quickly deactivated (Eq.\ref{ts}) and short parabolic dunes emerge (Fig.~3a). At large $\theta$ the amount of sand trapped by plants in the dune arms, which height scales as $V^{2/3}/L\sim V^{1/3}\,(\theta_c-\theta)$, is reduced. This leads to longer parabolic dunes whose `noses' can experience successive splits before being finally deactivated (Fig.~3b and~3c).

\begin{figure}[htb]
\includegraphics[width=0.45\textwidth]{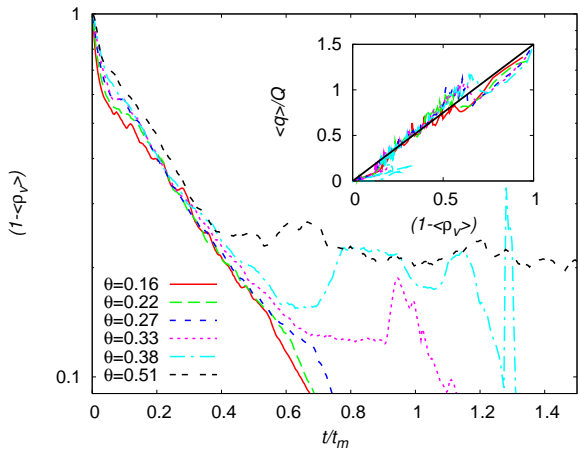}
\caption{The fraction of available sand surface $1-\langle \rho_v\rangle$, where $\langle \rho_v\rangle$ is the mean vegetation cover density, shows an exponential decay during the evolution of a vegetated barchan dune for small $\theta$ when time is normalized by $t_m$. Inset: proportionality between the mean sand flux $\langle q\rangle$, normalized by the saturated upwind value $Q$, and the fraction of active sand surface.}
\end{figure}

The dependence of $t_s$ on the properties of the initial barchan through $\theta$ and $t_m$ leads us to characterize the mobility or activity of a vegetated dune only based on general field conditions. In particular, the condition $\theta<\theta_c$ for a migrating barchan to become inactive explicitely reads
\begin{equation}
\label{qc}
Q < Q_c\equiv \theta_c\,V_v V^{1/3} \, ,
\end{equation}
which represents a size dependent upper limit for $Q$ below which inactivation takes place, and thus stresses the leading role of the wind strength in the inactivation process. It also defines a minimum vegetation growth velocity $V_{vc} =\theta_c^{-1} Q/V^{1/3}$, also a function on dune size, and dune volume $V_c = (\theta_c^{-1} Q/V_v)^3$ above which barchan dunes are stabilized. Furthermore, the dune inactivation time as explicit function of $V$ achieves a minimum for a dune volume $V_0 = (3/2)^3\,V_c$. Hence, dunes with volumes below $V_c$ will remain mobile, while dunes with volumes $V_0$ just above the minimum one will be the first to be stabilized. Large dunes however, will remain mobile for a time that scales with $t_m \propto V^{2/3}$. 

The dynamics of dune stabilization is also characterized by the evolution of the mean vegetation cover density $\langle \rho_v\rangle$ (Fig.~4) and, in a minor way, by the mean sand flux $\langle q\rangle$ over the dune, which we find to be proportional to the fraction of available sand surface $1-\langle \rho_v\rangle$ (Fig.~4, inset). For small $\theta$ the evolution follows the same trend, e.g. a rising vegetation cover and the resulting reduction of the mean sand flux \cite{lancaster98}, with $t_m$ as characteristic time. For larger $\theta$, the vegetation cover reaches a plateau that extends until the dune stabilizes at $t\approx t_s$, after which vegetation cover rises again due to the inactivation of the dune. This plateau extends indefinitely at $\theta=\theta_c$, when $t_s$ diverges, as a sign of actual sand mobility.

We have proposed a set of differential equations of motion describing the aeolian transport on vegetated granular surfaces including the growth and destruction of plants. Through them we calculate the morphological transition between active barchans and inactive parabolic dunes characterized by the divergence of the dune inactivation time $t_s$ and its scaling with the barchan migration time $t_m$.  We find that the fixation index $\theta$, a dimensionless combination of barchan dune size, wind strength and vegetation growth velocity, determines from the very beginning the final outcome of the competition between vegetation growth and aeolian surface mobility. These predictions well be directly tested in the field and have implications on the economy of semi-arid regions on coastal management and on global ecosystems.

We thank P.G.~Lind, V.~Schw\"ammle and H. Tsoar for useful discussions. This study was supported by the DFG, the Max Planck Preis and the Volkswagenstiftung.

\end{document}